\documentclass[conference, a4paper, 10pt]{IEEEtran}
\usepackage{geometry}
    
\usepackage{mathrsfs}
\usepackage{amssymb}
\usepackage[mathcal]{euscript}
\usepackage{cite}
\usepackage{graphicx}
\usepackage{psfrag}
\usepackage{subfigure}
\usepackage{url}
\usepackage{stfloats}
\usepackage{amsmath}
\usepackage{float}
\usepackage{array}

\usepackage{amsthm}
\usepackage{booktabs} %±ížñ
\usepackage{algorithm} %format of the algorithm
\usepackage{algorithmic} %format of the algorithm
\usepackage{color}
\usepackage{cases}
\usepackage{bm}
\usepackage{amsmath,amsfonts,amssymb,amsbsy,paralist,ifthen}

\usepackage{colortbl}
\usepackage{arydshln}
\usepackage{multirow}
\usepackage{multicol}

\usepackage{color}

\ifCLASSINFOpdf
\else
\fi
\begin{document}

\title{Heuristic Deep Reinforcement Learning for Phase Shift Optimization in  RIS-assisted Secure Satellite Communication Systems with RSMA}

\author{Tingnan Bao\IEEEauthorrefmark{2}, \textit{Senior Member, IEEE},  Melike Erol-Kantarci\IEEEauthorrefmark{2},\textit{ Senior Member, IEEE}\\
\IEEEauthorblockA{\IEEEauthorrefmark{2}\textit{School of Electrical Engineering and Computer Science, University of Ottawa, Ottawa, Canada}}
\IEEEauthorblockA{Emails:\{tbao, melike.erolkantarci\}@uottawa.ca}}

\maketitle

\begin{abstract}
%\boldmath
This paper presents a novel heuristic deep reinforcement learning (HDRL) framework designed to optimize reconfigurable intelligent surface (RIS) phase shifts in secure satellite communication systems utilizing rate splitting multiple access (RSMA). The proposed HDRL approach addresses the challenges of large action spaces inherent in deep reinforcement learning by integrating heuristic algorithms, thus improving exploration efficiency and leading to faster convergence toward optimal solutions. We validate the effectiveness of HDRL through comprehensive simulations, demonstrating its superiority over traditional algorithms, including random phase shift, greedy algorithm, exhaustive search, and Deep Q-Network (DQN), in terms of secure sum rate and computational efficiency. Additionally, we compare the performance of RSMA with non-orthogonal multiple access (NOMA), highlighting that RSMA, particularly when implemented with an increased number of RIS elements, significantly enhances secure communication performance. The results indicate that HDRL is a powerful tool for improving the security and reliability of RSMA satellite communication systems, offering a practical balance between performance and computational demands.

\end{abstract}

\begin{IEEEkeywords}
Heuristic deep reinforcement learning, RIS, RSMA, secure satellite communication, physical layer security
\end{IEEEkeywords}

\vspace{-0.1 in}
\section{Introduction}\label{intro_1}
The rapid growth of global communication demands, particularly with the advent of 6G networks, has led to an increasing need for efficient and secure satellite communication systems. Satellite communication, particularly involving low earth orbit (LEO) satellites, faces unique challenges such as long-distance transmission, high latency, and inherent eavesdropping vulnerabilities in wireless channels. Non-orthogonal multiple access (NOMA) has gained widespread adoption in this context due to its ability to improve spectral efficiency by enabling multiple users to share the same frequency resources through power domain multiplexing \cite{gao2020performance}. However, despite its capacity advantages, NOMA still faces difficulties in managing the resulting interference and ensuring robust security, particularly in the complex environments typical of satellite communication \cite{kodheli2020satellite}. These challenges have prompted researchers to explore alternative and complementary technologies that can offer better interference management and enhanced security.

To address these challenges, rate splitting multiple access (RSMA) has emerged as an alternative technology that offers improved interference management by dividing user signals into common and private components \cite{yin2022rate}. Most previous research work \cite{cui2023energy}, \cite{khan2023rate}, \cite{lee2023coordinated} has proven that RSMA can enhance the overall efficiency and robustness of satellite communication systems. 
However, it is noteworthy that satellite communication, being a form of wireless communication, is inherently vulnerable to eavesdropping. To mitigate this vulnerability and further strengthen the security of RSMA-based satellite systems, physical layer security (PLS) has gained attention to enhance the confidentiality and integrity of transmitted data \cite{lin2020secure}. By integrating PLS with RSMA, satellite communication systems can not only manage interference more effectively but also provide an additional layer of protection against eavesdropping.

Building upon these advancements, reconfigurable intelligent surfaces (RIS) can further enhance the performance of RSMA-based secure satellite communication systems by dynamically adjusting the phase shifts of reflected signals to improve overall signal strength \cite{jiang2023aerial,bao2021adep,bao2022performance}. However, optimizing RIS phase shifts in such systems presents significant challenges due to the large number of RIS elements involved. In \cite{lin2022refracting}, the authors proposed an alternating optimization (AO) scheme to jointly optimize beamforming and phase shifts of RIS in satellite relay networks. Similarly, in \cite{wang2022secure}, successive convex approximation methods and the S-Lemma were employed to optimize the phase shifts of RIS in secure RIS-assisted satellite transmissions. 

Despite these efforts, the primary drawback of AO schemes, as traditional methods for optimizing RIS phase shifts, is their tendency to converge to suboptimal local minima, particularly in complex, high-dimensional problems. This limitation has prompted the exploration of deep reinforcement learning (DRL) approaches, which offer more effective and globally optimal solutions. Although \cite{liu2023deep} introduced a double cascade correlation network (DCCN) to adjust RIS reflection coefficients and optimize unmanned aerial vehicle (UAV) trajectories in UAV-RIS-assisted cognitive secure non-terrestrial networks, the action space remains large, leading to potential inefficiencies. These limitations underscore the need for more efficient optimization strategies that can effectively manage the complexity of RIS in RSMA-based secure satellite communication systems.

Motivated by the above, we adapt a heuristic DRL (HDRL) in \cite{zhou2024heuristic} by 
integrating heuristic algorithms with DRL to efficiently optimize RIS phase shifts in secure satellite communication systems. 
The main contributions of this work are summarized as follows:

\begin{itemize}
    \item We adpat HDRL into optimizing RIS phase shifts in secure satellite communication systems employing the RSMA scheme.
    \item We demonstrate the effectiveness of HDRL in achieving higher secure sum rates and improved computational efficiency compared to traditional methods, including random phase shift, greedy, exhaustive search, and Deep Q-network (DQN).
    \item We present a comparative analysis of RSMA and NOMA within the secure satellite communication network, emphasizing the superior performance of RSMA, particularly when combined with a greater number of RIS elements.
   % \item We validate the proposed framework through simulations, showing its benefits in enhancing security and performance in satellite communication networks.
\end{itemize}

%The rest of this paper is organized as follows: Section II describes the system model, including the RSMA and RIS configurations. Section III introduces the proposed HDRL framework, detailing the state space, action space, and learning algorithm. Section IV presents the simulation results, comparing the performance of HDRL with existing methods. Finally, Section V concludes the paper.
\vspace{-0.1 in}
\section{System Model} \label{sys_model}
%\subsection{Network Model}
\begin{figure}[t!]
\centering
\includegraphics[width=0.475\textwidth]{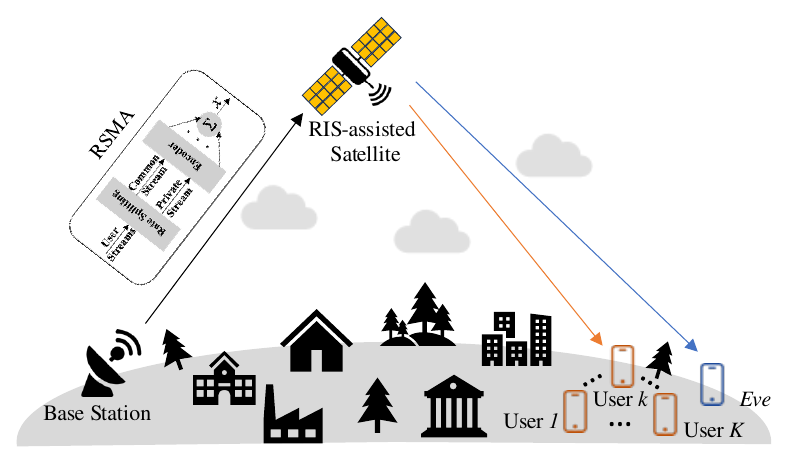}
\caption{Model of an RIS-assisted secure satellite communication system with the RSMA scheme.}
\label{fig_sys}
\vspace{-0.2 in}
\end{figure}

We consider an RIS-assisted multi-user downlink RSMA secure satellite communication system, as illustrated in Fig. \ref{fig_sys}. In this system, an $M$-antenna base station (BS) simultaneously transmits messages to $K$ single-antenna users, aided by a satellite equipped with an $L$-element passive RIS, in the presence of an eavesdropper referred to as Eve ($e$). The direct link is assumed to be obstructed due to the large distance between the BS and the users/Eve. 

The RSMA scheme employed at the BS divides the transmitted message $s$ into two parts: the common part, which is encoded into a shared codebook to generate a single data stream, denoted as $s_c$, and the private part, which is encoded into a private codebook specific to the $k$-th user, denoted as $s_k$. We assume the transmitted message has unit power, $\mathbb{E}[ss^H] = 1 $. The passive RIS, comprising $L$ reflective elements, has a phase shift matrix represented by $\boldsymbol{\Phi} = \mathrm{diag}(\beta_1 e^{j \theta_1}, \beta_2 e^{j \theta_2}, \cdots, \beta_L e^{j \theta_L}) \in \mathbb{C}^{L \times L}$, where $\beta_l$ and $\theta_l$ are the amplitude and phase shift for the $l$-th reflecting element of the RIS. For simplicity, we assume $\beta_l = 1$ for $l \in \{1, \cdots, L\}$, and the phase shift $\theta_l \in \left\{0, \frac{2\pi}{2^\mu}, \cdots, \frac{2\pi (2^\mu - 1)}{2^\mu} \right\}$ with a resolution of $\mu$.

\subsection{Channel Model}
The channel between the BS and users/Eve is subject to both large-scale and small-scale fading effects. Considering that the RIS is equiped on the satellite, the large-scale fading for the BS-RIS and RIS-User $k$/Eve links can be modeled by the free-space path-loss model, expressed as
\begin{align}
    L(d_{S,R})= G_S \left( \frac{c}{4 \pi f_c} \right)^2 d_{S,R}^{-\alpha_{S,R}},
\end{align}
and
\begin{align}
        L(d_{R,B})= G_B \left( \frac{c}{4 \pi f_c} \right)^2 d_{R,B}^{-\alpha_{R,B}},
\end{align}
respectively, where $d_{S,R}$ and $d_{R,B}$ represent the distances from the BS to the RIS, and from the RIS to the target user $B$, where $B \in \{k:$ User $k$, $e:$ Eve$\}$, $G_S$ and $G_B$ denote the antenna gains of the BS and the target user $B$, $c$ is the speed of light, $f_c$ is the Ku carrier frequency, and $\alpha_{S,R}$ and $\alpha_{R,B}$ are the path loss exponents for the BS-RIS and RIS-B links, respectively. 

Under the complex environment in this secure system, both BS-RIS and RIS-B links follow the Rician fading distrubition \cite{peng2022performance}, which can be expressed by
l
respectively, where $\textbf{G} \in \mathbb{C}^{L \times M}$ denotes the channel matrix between the BS and the RIS, $\textbf{h}_B \in \mathbb{C}^{1 \times L}$ denotes the channel vector between the RIS and the target user $B$, $\kappa$ and $\mu_\kappa$ are the Rician factors for the BS-RIS and RIS-B links. The matrix $\widetilde{\textbf{G}} \in \mathbb{C}^{L \times M}$ and the vector $\widetilde{\textbf{h}}_B \in \mathbb{C}^{1 \times L}$ represent the non-line-of-sight (NLoS) parts of the channels. The matrix $\overline{\textbf{G}} \in \mathbb{C}^{L \times M}$ and the vector $\overline{\textbf{h}}_B \in \mathbb{C}^{1 \times L}$ represent the LoS parts of the channels, respectively. 

\subsection{RSMA-based Transmission}
In this system, the BS transmits messages to users using the RSMA scheme. Let $p_c$ and $p_k$ denote the transmit power of the common and private streams, respectively. A split factor $\alpha \in (0,1]$ is introduced for the allocation of total transmit power $P_t$ \cite{schroder2023comparison}. Accordingly, the common stream $s_c$ is allocated $p_c = \alpha P_t$, while the remaining private streams share $\sum_{k=1}^K p_k = (1-\alpha) P_t$. For each private stream $s_k$, we assume uniform power allocation, meaning that $p_k = (1-\alpha)P_t/K$. The resulting transmit signal at the BS can be expressed as
\begin{align}
   \textbf{x} = \sqrt{p_{\textrm{c}}} \textbf{w}_{\rm{c}} s_{\textrm{c}} + \sum_{k=1}^{K} \sqrt{p_{\textrm{k}}} \textbf{w}_{\rm{k}} s_{\textrm{k}},
 \end{align}
where $\textbf{w}_{\rm{c}} \in \mathbb{C}^{M \times 1} $ and $\textbf{w}_{\rm{k}} \in \mathbb{C}^{M \times 1} $ are the unit-norm precoding vectors for the common stream $s_c$ and the $k$-th private stream $s_k$, respectively.
Thus, the power of the transmitted signal from the multi-antenna BS is given by 
\begin{align} \label{x2}
   \mathbb{E} \left[|\textbf{x}|^2 \right] = \text{tr}(\textbf{P}\textbf{W}^H\textbf{W}) = p_c + \sum_{k=1}^K p_k = P_t,
\end{align}
where $\textbf{W} \triangleq [\textbf{w}_{\rm{c}}, \textbf{w}_{\rm{1}}, \textbf{w}_{\rm{2}}, 
\cdots, 
\textbf{w}_{\rm{K}}] \in \mathbb{C}^{M \times (K+1)}$ and $\textbf{P} \triangleq \text{diag} [{p}_{\rm{c}}, {p}_{\rm{1}}, {p}_{\rm{2}}, 
\cdots, 
{p}_{\rm{K}}] \in \mathbb{R}^{K \times (K+1)}$. 
Then, the received signal at user $k$ can be given by
\begin{align}
    y_k= \sqrt{L(d_{S,R}) L(d_{R,k}) g_s } \textbf{h}_k \boldsymbol{\Phi} \textbf{Gx}  + N_0,
\end{align}
where $g_s$ is the spread spectrum gain and $N_0$ is the additive white Gaussian noise (AWGN) with a variance of $\sigma^2$. The received signal-to-interference-plus-noise ratio (SINR) of the common parts at the $k$-th user can be expressed as follows
\begin{align}
    \gamma_k^c= \frac{ \rho_k p_c | \textbf{h}_k \boldsymbol{\Phi} \textbf{G} \textbf{w}_c |^2  }{ \sum_{j=1}^K \rho_k p_j  | \textbf{h}_j \boldsymbol{\Phi} \textbf{G} \textbf{w}_j |^2  +\sigma^2 },
\end{align}
where $\rho_k = L(d_{S,R}) L(d_{R,k}) g_s$. After successfully decoding $s_c$, its contribution to $y_k$ is eliminated using SIC, allowing for the decoding of the private parts. As such, the received SINR of the private parts for the $k$-th user can be given by
\begin{align}
        \gamma_k^p= \frac{ \rho_k p_k | \textbf{h}_k \boldsymbol{\Phi} \textbf{G} \textbf{w}_k |^2  }{ \sum_{j=1, j \neq k}^K \rho_k p_j  | \textbf{h}_j \boldsymbol{\Phi} \textbf{G} \textbf{w}_j |^2  +\sigma^2 }.
\end{align}
Then, the achievable rates of the common and private parts at the $k$-th user are
\begin{align}
      C_{k}^c= B_w \log_2 (1+ \gamma_k^c) \:\: \text{and} \:\: C_{k}^p= B_w \log_2 (1+ \gamma_k^p) ,
\end{align}
respectively, where $B_w$ is the available bandwidth. 

Similarly, the eavesdropping SINR for both the common and private parts, assuming that the statistical CSI is available to Eve \cite{pei2024secrecy}, can be expressed as
\begin{align}
     \gamma_e^c= \frac{ \rho_e p_c | \textbf{h}_e \boldsymbol{\Phi} \textbf{G} \textbf{w}_c |^2  }{ \sum_{j=1}^K \rho_e p_j  | \textbf{h}_e \boldsymbol{\Phi} \textbf{G} \textbf{w}_j |^2  +\sigma^2 },
\end{align}
and
\begin{align}
     \gamma_e^p= \frac{ \rho_e p_k | \textbf{h}_e \boldsymbol{\Phi} \textbf{G} \textbf{w}_k |^2  }{ \sum_{j=1, j \neq k}^K \rho_e p_j  | \textbf{h}_e \boldsymbol{\Phi} \textbf{G} \textbf{w}_j |^2  +\sigma^2 },
\end{align}
respectively, where $\rho_e = L(d_{S,R}) L(d_{R,e}) g_s$. Then, the achievable rates of the common and private parts at Eve are 
\begin{align}
      C_{e}^c= B_w \log_2 (1+ \gamma_e^c) \:\: \text{and} \:\: C_{e}^p= B_w \log_2 (1+ \gamma_e^p) ,
\end{align}
respectively. Thus, the secrecy rates of the common and private parts at the $k$-th user can be expressed as
\begin{align}
    R_{k}^c= \text{max}[C_{k}^c - C_{e}^c,0] \:\: \text{and} \:\:  R_{k}^p= \text{max}[C_{k}^p - C_{e}^p,0]  .  
\end{align}
The overall secrecy rate at the $k$-th user is given by $R_k = R_{k}^c + R_{k}^p $. 

\subsection{Problem Formulation}
Given that uniform power allocation is considered, the values of $\textbf{w}_c$ and $\textbf{w}_k$ are deterministic. Therefore, the optimization focuses on maximizing the secure sum rate for all users in the presence of Eve by adjusting the phase shifts at the RIS. The problem formulation is as follows:
\begin{equation} \label{optimatization}
\begin{aligned}
      \max_{\Phi} \quad & \sum_{k=1}^K R_k \\
\textrm{s.t.} \quad & C1: (\ref{x2})\\
& C2: || \textbf{w}_c ||^2 =1, || \textbf{w}_k ||^2 =1, \forall k= 1, \dots, K \\   
& C3: |\theta_l|^2 = 1 , \forall l= 1, \dots, L  \\  
& C4: \theta_l \in \left\{0, \frac{2\pi}{2^\mu}, \dots, \frac{2\pi (2^\mu - 1)}{2^\mu} \right\}, \forall l= 1, \dots, L  \\
\end{aligned}
\end{equation}
where \( C1 \) represents the power constraint, \( C2 \) ensures that the normalization constraint is satisfied, which is crucial for maintaining a consistent power level for the RSMA scheme, and \( C3 \) imposes the condition that the magnitude of each phase shift \( \theta_l \) must be 1. In other words, each element in the RIS phase shift matrix acts as a pure phase shifter. \( C4 \) reflects that the phase shifts are limited to discrete levels determined by the resolution \( \mu \). These constraints, combined with the non-linear and non-convex nature of the objective function in (\ref{optimatization}), make the optimization problem challenging to solve. To address this, we propose a novel HDRL approach in the following section. 
 \vspace{-0.1 in}
\section{Heuristic Deep Reinforcement Learning}
\label{hdrl_model}
\begin{figure}[t!]
\centering
\includegraphics[width=0.475\textwidth]{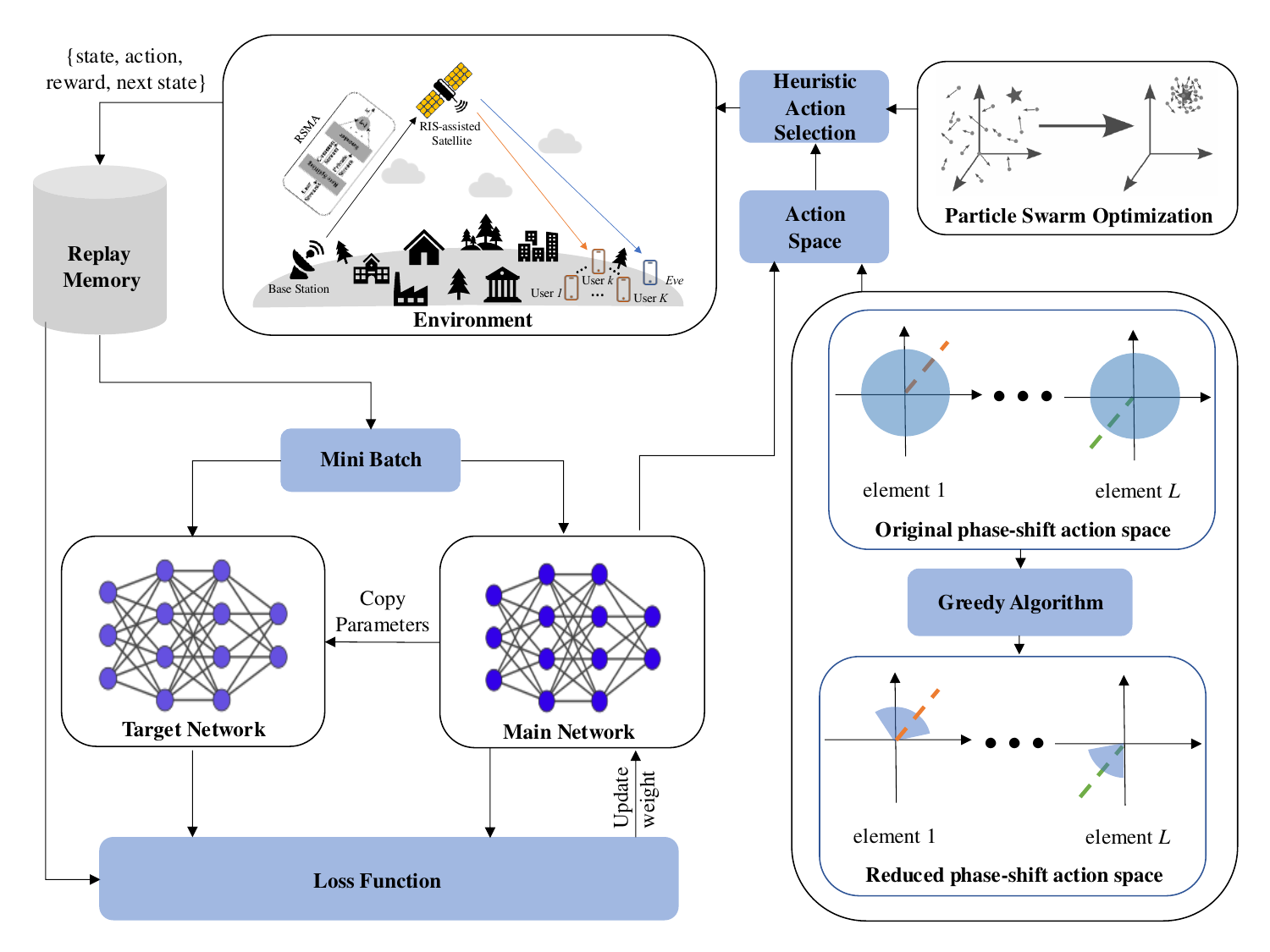}
\caption{Framework of HDRL for RIS-assisted RSMA secure satellite communciatons.}
\label{fig_hdrl}
\vspace{-0.2 in}
\end{figure}
To efficiently tackle this issue, we transformed the problem into a Markov Decision Process (MDP) framework. While DRL can be utilized to optimize RIS phase shifts \cite{wu2024deep,ngo2023multi,wu2023joint}, the large action spaces often hinder exploration efficiency and may lead to suboptimal solutions. To overcome this challenge, we propose the HDRL method, which integrates heuristic algorithms into DRL, as shown in Fig. (\ref{fig_hdrl}). This approach enhances exploration efficiency and steers the results toward a global optimum. Thus, to fully explain how HDRL operates within the context of an RIS-assisted satellite communication system, we can begin by understanding the traditional DRL framework and then explain the modifications introduced by HDRL. The following section will delve into the concepts of state, action, and reward, and how HDRL optimizes these aspects.
\subsection{State Space}
In the RIS-assisted satellite communication system that employs the RSMA scheme, the state ${s}_t$ at time $t$, which includes the current conditions of the system, can be expressed as follows:
\begin{align}
   {s}_t =  \{ \boldsymbol{\Phi}_t, \textbf{G}_t, \textbf{h}_k (t),   \textbf{h}_e (t) \},
\end{align}
where $\boldsymbol{\Phi}_t$ is the phase shift matrix of the RIS, $\textbf{G}_t$ is the channel matrix from the BS to the RIS, 
$\textbf{h}_k (t)$ is the channel vector from the RIS to user $k$, and $\textbf{h}_e (t)$ is the channel vector from the RIS to Eve. 
\vspace{-0.1 in}
\subsection{Action Space}
%In the context of optimizing the RIS phase shifts for a satellite communication system using HDRL, the action space is a critical component that directly influences the efficiency and effectiveness of the learning process. Here, we detail the process of defining and reducing the action space, incorporating both greedy algorithms and heuristic methods to enhance the DRL framework
Since the action space is a critical component that directly influences the efficiency and effectiveness of the learning process, we detail the process of defining and reducing the action space. This incorporates both greedy algorithms and heuristic methods to enhance the DRL framework. The action space in a traditional DRL can be expressed by
\begin{align}
    \mathcal{A} = & \bigg\{ \boldsymbol{\Phi}_t | \boldsymbol{\Phi}_t = diag (e^{j\theta_1}, e^{j\theta_2}, \cdots, e^{j\theta_L} )\bigg\}, %\nonumber\\
  % & \theta_l \in \left\{0, \frac{2\pi}{2^\mu}, \cdots, \frac{2\pi (2^\mu - 1)}{2^\mu} \right\} \bigg\}. 
\end{align}
To make the exploration of this large action space more feasible, we first apply a greedy algorithm, which operates on an element-by-element basis. This algrithom selects the phase shift for each RIS element that maximizes the secure sum rate $R_t$. This can be expressed by
\begin{align}
    \theta_l^{\text{greedy}}= \arg \max_{\theta_l}  \mathbb{E} [R_t|s_t, \theta_t].
\end{align}
This element-wise optimization is performed sequentially for all $L$ elements, producing an initial configuration $\boldsymbol{\Phi}^{\text{greedy}}$. Then, we conduct a frequency analysis to identify the phase shifts that are most frequently selected for each RIS element. These frequent phase shifts are considered more promising and form the basis for the reduced action space $\mathcal{A}{'}$, which can be defined as
\begin{align}
    \mathcal{A}{'} = \{ \boldsymbol{\Phi}_t | \theta_l \in \{ \theta_{l,1}, \theta_{l,2}, \cdots, \theta_{l,n_l}  \} , 
\end{align}
where $\theta_{l,1}, \theta_{l,2}, \cdots, \theta_{l,n_l}$ are the most frequently selected phase shifts from $\boldsymbol{\Phi}^{\text{greedy}}$ for the $l$-th element. Thus, in our proposed HDRL framework, the reduced action space $ \mathcal{A}{'}$ becomes the new action space for DRL. This reduced space significantly improves exploration efficiency because the agent now only needs to explore a smaller, more promising set of phase-shift configurations. Furthermore, meta-heuristic algorithms, such as particle swarm optimization (PSO) or genetic algorithms (GA),  can be used to further refine the action selection process within this reduced space. As such, the action selection in HDRL is given by
\begin{align}
    a_t = \arg \max_{\boldsymbol{\Phi}_t \in \mathcal{A}{'}} \mathbb{E} [R_t|s_t, \boldsymbol{\Phi}_t ].
\end{align}
\vspace{-0.1 in}
\subsection{Reward Function}
The reward $R_t$ at time $t$, which reflects the quality of the action taken by the agent, can be expressed by
\begin{align}
    R_t = \sum_{k=1}^K R_k.
\end{align}
\vspace{-0.1 in}
\subsection{Learning Algorithm}
The learning algorithm in our HDRL framework is designed to optimize RIS phase shifts in satellite communication systems by deploying a  DQN with heuristic methods. This network takes the current state $s_t$ as input and outputs the Q-values $Q(s_t, a_t; \varrho)$ for all possible actions $a_t$. To update the network, loss function can be given by
\begin{align} \label{lossfunction}
    L( \varrho) = &\frac{1}{N_B} \sum_{j=1}^{N_B} \bigg(  R_j + \gamma_\text{HDRL} \max_{a' \in \mathcal{A}{'}} Q(s_{t+1}, a';  \varrho^{-} ) \nonumber\\
    & Q(s_j,a_j; \varrho )
       \bigg)^2,
\end{align}
where $N_B$ is the number of samples in the mini-batch, $\gamma_\text{HDRL}$ is the discount factor, $a'$ is the set of all possible actions that can be considered at the next time step when the agent is at the state $s_{t+1}$, and $\varrho$ and $\varrho^{-}$ are the parameters of training and target networks, respectively. To enhance exploration efficiency, HDRL employs heuristic methods, beginning with a greedy algorithm that reduces the action space by selecting phase shifts that maximize immediate rewards for each RIS element. The reduced action space $\mathcal{A}{'}$ is then utilized in the Q-learning process, focusing on the most promising actions. The Q-value update for each step, performed using the Bellman equation, can be expressed by
\begin{align}
    Q(s_t, a_t) \leftarrow & Q(s_t, a_t) + \alpha_\text{HDRL} \bigg[ R_t + \gamma_\text{HDRL}\nonumber\\
   &
     \max_{a' \in \mathcal{A}{'}} Q(s_{t+1}, a'; \varrho^{-} )  -Q(s_t,a_t; \varrho) \bigg],
\end{align}
where $\alpha_\text{HDRL}$ is the learning rate. This update ensures that the agent improves its decisions by continuously adjusting the Q-function to reflect the best possible actions. Furthermore, the learning algorithm also manages the exploration-exploitation trade-off using an 
$\epsilon$-greedy strategy. Thus, by reducing the action space and focusing on the most promising actions, HDRL effectively narrows down the search space, making it more feasible to explore globally optimal solutions. The proposed approach’s pseudo-code is outlined
and depicted in Algorithm 1.

%Finally, the policy $\pi_\text{HDRL}$  is derived from the Q-function by selecting the action that maximizes $Q(s_t,a_t)$ within the reduced action space $\mathcal{A}{'}$, can be given by
%\begin{align}
%   \pi_\text{HDRL} (s_t) = \arg \max_{a_t \in \mathcal{A}{'}  } Q(s_t,a_t). 
%\end{align}
%This results in a more focused and efficient policy that effectively optimizes the RIS phase shifts in satellite communication systems.

\begin{algorithm} \label{algorithm}
\caption{HDRL for RIS Phase Shifts Optimization}
\begin{algorithmic}[1]
\STATE Initial state $s_0$, discount factor $\gamma_\text{HDRL}$, learning rate $\alpha_\text{HDRL}$, max episodes $N$, max steps per episode $T$, replay memory pool $D$, batch size $N_B$, and exploration probability $\epsilon$
\STATE Initialize training and target network parameters $\varrho$, $\varrho^-$, respectively, and reduced action space $\mathcal{A}'$ using a greedy algorithm and heuristics
\STATE Set current state $s \gets s_0$

\FOR{episode $n = 1$ to $N$}
    \STATE Set initial state $s \gets s_0$
    \FOR{each step $t = 1$ to $T$}
        \STATE With $\epsilon$, select random action $a_t$ from $\mathcal{A}'$
        \STATE Otherwise, select $a_t = \arg\max_{a \in \mathcal{A}'} Q(s_t, a; \varrho)$
        \STATE Execute $a_t$, observe $R_t$ and $s_{t+1}$
        \STATE Store $(s_t, a_t, R_t, s_{t+1})$ in $D$
        \STATE Sample mini-batch of $(s_j, a_j, R_j, s_{j+1})$ from $D$
        \FOR{each sampled transition}
            \STATE Perform gradient descent to minimize loss utilizing Eq. (\ref{lossfunction}).
        \ENDFOR
        \STATE Update state $s_t \gets s_{t+1}$
    \ENDFOR
    \STATE Periodically update $\varrho^-$ with weights from $\varrho$
    \STATE Decay $\epsilon$ after each episode
\ENDFOR
\end{algorithmic}
\end{algorithm}
\vspace{-0.1 in}
\section{Simulation Results}
In this section, we aim to demonstrate the performance of the proposed HDRL approach for optimizing RIS phase shifts in secure satellite communication systems utilizing the RSMA scheme. The environment settings and model parameters used in these simulations are detailed in Table \ref{tb_1}. In our system, a multi-antenna BS transmits signals to five single-antenna users, randomly located at distances between 300 km and 500 km from the RIS, with a transmit power split factor of 0.3 under the RSMA scheme, in the presence of one eavesdropper. The system is further enhanced by varying the number of RIS elements, ranging from 20 to 60, within the satellite communication environment. 

\begin{table}[t!] %table ÀïÃæÒ²¿ÉÒÔÇ¶Ì×tabular,Ö»ÓÐtabular ÊÇ²»ÄÜŒÓ±êÌâµÄ
\renewcommand{\arraystretch}{1.5}
\centering  %±ížñŸÓÖÐ
\caption{System Parameters} %±ížñ±êÌâ
 \begin{tabular}{m{4.4cm}|m{0.9cm}|m{2.2cm}}  %ÓÒ¶ÔÆë
     \hline
     \hline
      Parameters & Notations & Values \\ %[0.05ex]  %ÔöŒÓÐÐ¿í
       \hline
       Number of user
 & $K$ & 5 \\
        \cdashline{1-3}[0.8pt/2pt]
       Number of antenna at BS
 & $M$ & 3 \\
       \cdashline{1-3}[0.8pt/2pt]
       Number of RIS elements
 & $L$ & [20, 30, 40, 50, 60] \\
       \cdashline{1-3}[0.8pt/2pt]
       RIS phase shift resolution & $\mu$ & 2 \\
       \cdashline{1-3}[0.8pt/2pt]
       Number of Eve
 & $e$ & 1  \\
       \cdashline{1-3}[0.8pt/2pt]
        System bandwidth
 & $B_w$ & 20 MHz \\
       \cdashline{1-3}[0.8pt/2pt]
       Noise power spectral density
 & $\sigma^2$ & -96 dBm \\
       \cdashline{1-3}[0.8pt/2pt]
       Distance from the BS to the RIS
 & $d_{S,R}$ & 300 Km \\
       \cdashline{1-3}[0.8pt/2pt]
       Distance from the RIS to user & $d_{R,k}$ & [300, 500] Km \\
       \cdashline{1-3}[0.8pt/2pt]
       Distance from the RIS to Eve & $d_{R,e}$ & 450 Km \\
       \cdashline{1-3}[0.8pt/2pt]
       Transmit power at BS & $P_t$ & 70 dBm \\
       \cdashline{1-3}[0.8pt/2pt]
       Split factor of RSMA scheme& $\alpha$ & 0.3 \\
       \cdashline{1-3}[0.8pt/2pt]
       Anntenna gain at the BS& $G_S$ & 40 dBi \\
       \cdashline{1-3}[0.8pt/2pt]
       Anntenna gain at the target user B & $G_B$ & 20 dBi \\
       \cdashline{1-3}[0.8pt/2pt]
        Ku carrier frequency & $f_c$ & 14 GHz \\
        \cdashline{1-3}[0.8pt/2pt] 
        Speed of light & $c$ & $3.00 \times 10^8$ m/s \\
        \cdashline{1-3}[0.8pt/2pt]
       Spread spectrum gain & $g_s$ & 1000 \\
        \cdashline{1-3}[0.8pt/2pt]
        \multirow{2}{4.4cm}{Path-loss expontent and Rician factor for the BS-RIS link} & $\alpha_{S,R}$ & 2.0  \\
       \cdashline{2-3}[0.8pt/2pt]
        & $\kappa$ & 10 dB \\
       \cdashline{1-3}[0.8pt/2pt]
        \multirow{2}{4.4cm}{Path-loss expontent and Rician factor for the RIS-B link} & $\alpha_{S,B}$ & 2.5  \\
       \cdashline{2-3}[0.8pt/2pt]
        & $\mu_\kappa$ & 12 dB \\
       \cdashline{1-3}[0.8pt/2pt]
       \hline
       \hline
   \end{tabular}
   \label{tb_1}
   \vspace{-0.2 in}
\end{table}

\begin{table}[t!] 
\renewcommand{\arraystretch}{1.35}
\centering  
\caption{Hyperparameters for both DRL and HDRL } \label{tab_2}
\begin{tabular}{m{5.5cm}|m{2.4cm}}

\hline
\hline
Parameter & Value \\
	   
\hline
Number of neurons in each layer
 & (30,50,80) \\
\cdashline{1-2}[0.8pt/1pt]
Activation function &
ReLU \\ 
\cdashline{1-2}[0.8pt/1pt]
Optimizer &
Adam \\
\cdashline{1-2}[0.8pt/1pt]
Learning rate &
0.008 \\
\cdashline{1-2}[0.8pt/1pt]
Discount rate &
0.9 \\
\cdashline{1-2}[0.8pt/1pt]
Batch size &
10 \\
\cdashline{1-2}[0.8pt/1pt]
Experience Replay Pool &
1000 \\
\cdashline{1-2}[0.8pt/1pt]
Episode &
100 \\
\hline
\hline
\end{tabular}
\vspace{-0.2 in}
\end{table}

To verify the effectiveness of our proposed HDRL algorithm, we selected four benchmark optimization algorithms: random phase shift, greedy algorithm, exhaustive search, and DQN. These benchmarks provide a comprehensive comparison across different approaches to RIS phase shift optimization. The random phase shift serves as a non-optimized baseline, while the greedy algorithm offers an efficient, heuristic-based approach that may not achieve globally optimal results. Exhaustive search evaluates all possible configurations to find an optimal solution but is computationally expensive. The DQN algorithm is included for direct comparison with HDRL, with the hyperparameters for both detailed in Table II.

\begin{figure}[t!]
\centering
\includegraphics[height=5cm,width=6cm]{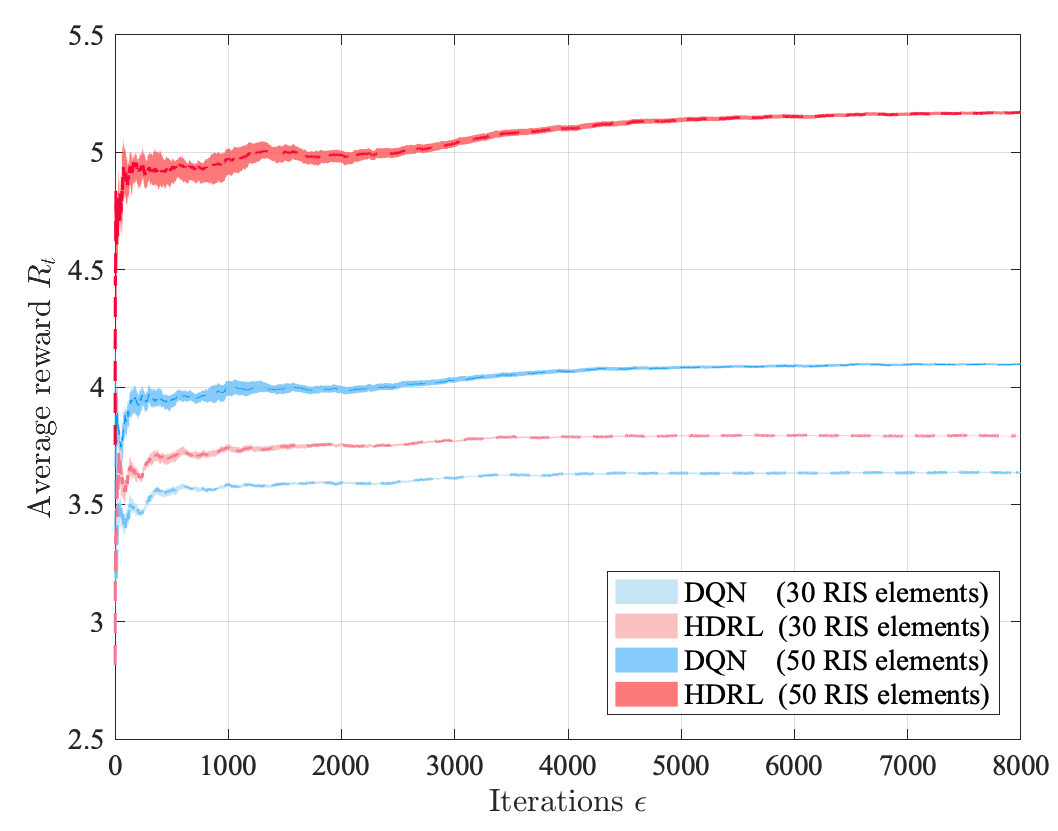}
\vspace{-0.2 in}
\caption{Average reward convergence vs. iterations for DQN and HDRL.}
\label{fig_reward_algri}
\vspace{-0.2 in}
\end{figure}

%Fig. \ref{fig_reward_algri} illustrates the convergence of average reward over iterations for both DQN and HDRL with different numbers of RIS elements. It can be observed that both algorithms exhibit an initial rise in average reward before reaching a stable state, indicating successful convergence for both DQN and HDRL. Notably, the HDRL consistently outperforms DQN across the same number of RIS elements, achieving higher average rewards. This improvement can be attributed to action pre-selection of HDRL using greedy and heuristic algorithms, which enhances exploration efficiency. Additionally, for both DQN and HDRL, the average reward is higher when 50 RIS elements are employed compared to 30 RIS elements. This suggests that utilizing a greater number of RIS elements in secure satellite communications can lead to improved system performance as expected.

Fig. \ref{fig_reward_algri} shows the convergence of the average reward over iterations for both DQN and HDRL with different numbers of RIS elements. Both algorithms successfully converge, but HDRL consistently achieves higher rewards due to its action pre-selection using greedy and heuristic algorithms, which enhances exploration efficiency. Additionally, increasing the number of RIS elements from 30 to 50 results in higher average rewards for both algorithms, indicating improved system performance.

\begin{figure}[t!]
\centering
\includegraphics[height=5cm,width=6cm]{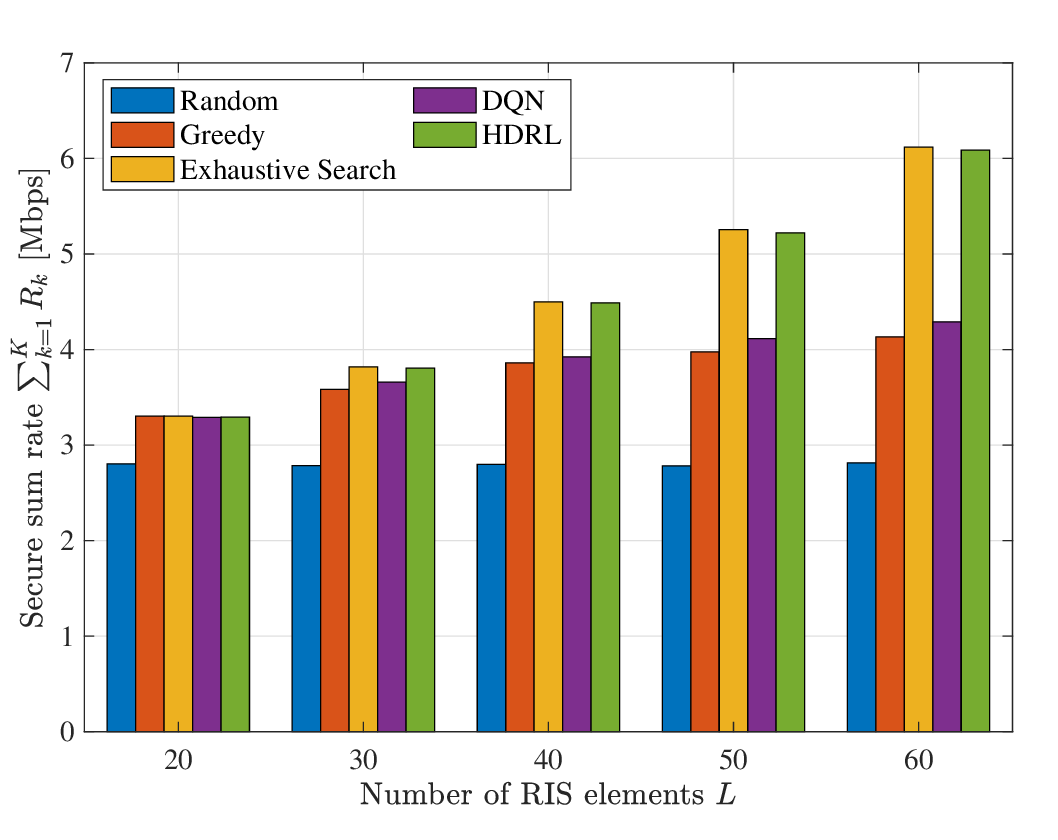}
\vspace{-0.2 in}
\caption{Secure sum rate vs. number of RIS elements for various algorithms.}
\label{fig_securesumrate}
\vspace{-0.2 in}
\end{figure}

Fig. \ref{fig_securesumrate} shows the secure sum rate as a function of the number of RIS elements for various algorithms. At lower numbers of RIS elements, such as 20, all algorithms except for the random phase shift approach perform relatively similarly. This is because the action space is smaller and less complex, making it easier for these algorithms to explore and find near-optimal solutions. As the number of RIS elements increases, HDRL consistently achieves the highest secure sum rate across all configurations, closely followed by the exhaustive search. DQN also performs well but lags behind HDRL and exhaustive search, particularly as the action space becomes larger with more RIS elements.

\begin{figure}[t!]
\centering
\includegraphics[height=5cm,width=6cm]{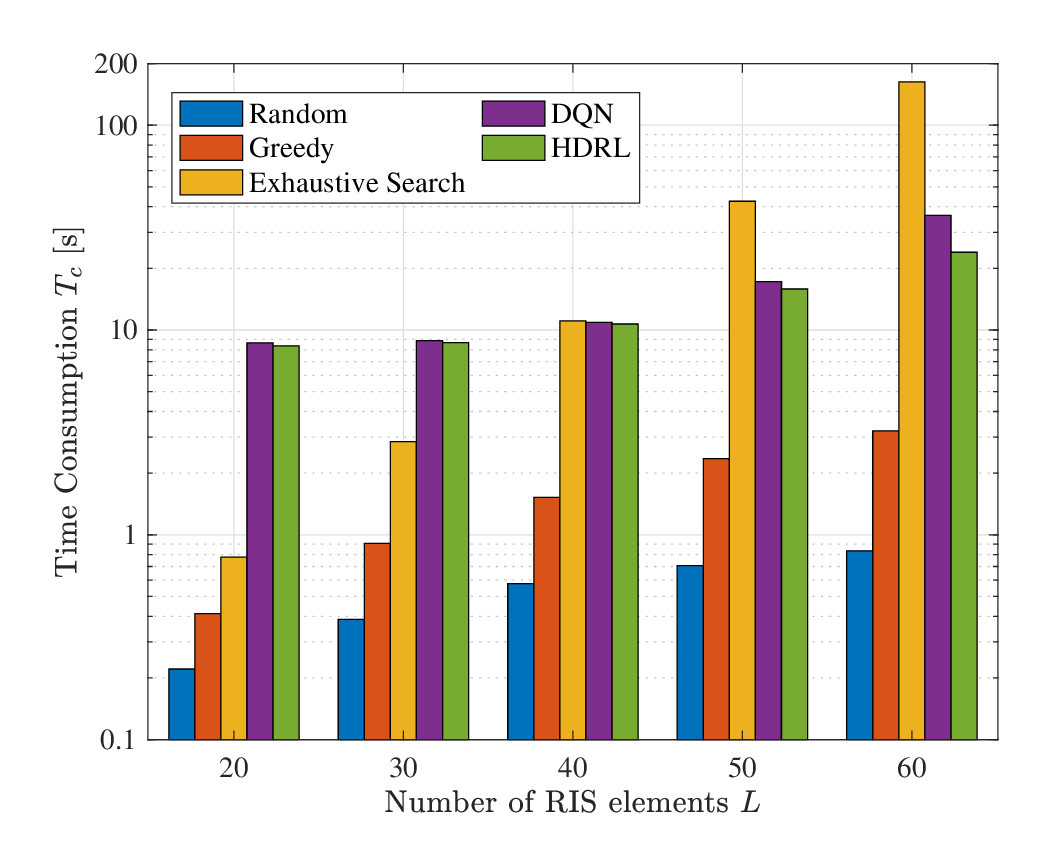}
\vspace{-0.2 in}
\caption{Time consumption vs. number of RIS elements for various algorithms.}
\label{fig_time}
\vspace{-0.2 in}
\end{figure}

Fig. \ref{fig_time} shows the time consumption as a function of the number of RIS elements for various algorithms. The random phase shift and greedy algorithms exhibit the lowest time consumption but lead to poorer system performance, as shown in Fig. \ref{fig_securesumrate}. The deterioration in
performance,as the number of RIS elements increases, is due to their lack of optimization and short-sighted decision-making. The exhaustive search, as expected, incurs the highest time consumption, especially at higher numbers of RIS elements (50 and 60 elements). In contrast, DQN and HDRL are more time-efficient than exhaustive search. However, the large action space of DQN causes a significant increase in computation time when the RIS elements reach 60. HDRL, on the other hand, maintains optimal performance with the least time consumption by leveraging a reduced action space. Thus, proposed HDRL delivers high performance with moderate computational demands.

\begin{figure}[t!]
\centering
\includegraphics[height=5cm,width=6cm]{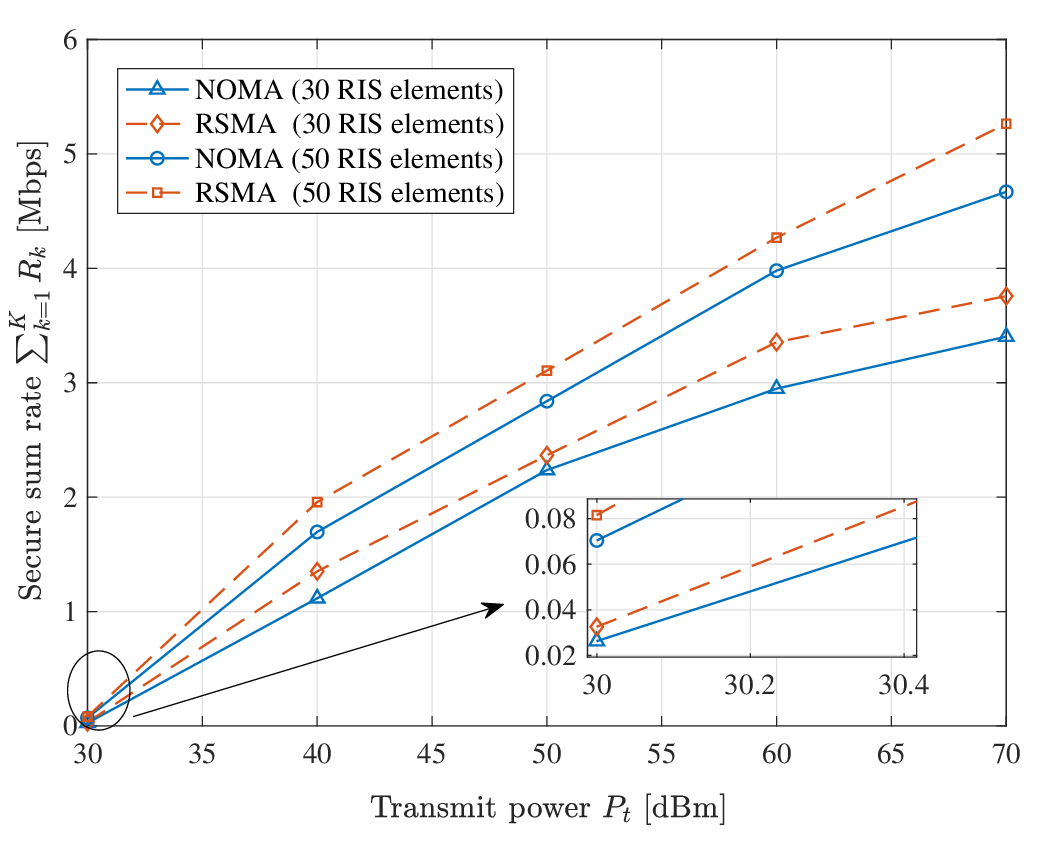}
\vspace{-0.2 in}
\caption{Secure sum rate vs. transmit power for RSMA and NOMA.}
\label{fig_power}
\vspace{-0.2 in}
\end{figure}

Additionally, to evaluate the effectiveness of the RSMA scheme, we also compared its performance against the NOMA benchmark. Fig. \ref{fig_power} shows the secure sum rate as a function of transmit power for both RSMA and NOMA schemes with 30 and 50 RIS elements. As transmit power increases from 30 dBm to 70 dBm, the secure sum rate improves for both schemes, with RSMA consistently outperforming NOMA at all power levels. Additionally, the secure sum rate is higher when 50 RIS elements are used compared to 30, indicating that increasing the number of RIS elements enhances system performance. The inset zooms in on the lower power range around 30 dBm, showing that RSMA maintains an advantage over NOMA even at lower power levels. Overall, the figure highlights RSMA's superiority in secure communication, particularly with more RIS elements and higher transmit power.

\vspace{-0.1 in}
 \section{Conclusion}
In this paper, we adpated a HDRL into optimizing RIS phase shifts in secure satellite communication systems, utilizing the RSMA scheme. Our approach effectively addresses the challenges posed by large action spaces, enhancing exploration efficiency and converging toward optimal solutions. Through comprehensive simulations, we have demonstrated that HDRL significantly outperforms traditional DQN and other benchmark algorithms, such as random phase shift, greedy algorithm, and exhaustive search, in terms of secure sum rate and computational efficiency. Additionally, our analysis confirmed that RSMA, especially when combined with an increased number of RIS elements, provides substantial gains over NOMA in secure communication performance. Overall, the proposed HDRL framework offers a robust and efficient solution to improve the security and reliability of RSMA satellite communication systems. 
\vspace{-0.3cm}
\section*{Acknowledgment}
This work has been supported by NSERC Canada Research Chairs program.
\vspace{-3mm}
\bibliographystyle{IEEEtran}
\bibliography{HDRL_RIS_Satellite}

\begin{thebibliography}{10}
\providecommand{\url}[1]{#1}
\csname url@samestyle\endcsname
\providecommand{\newblock}{\relax}
\providecommand{\bibinfo}[2]{#2}
\providecommand{\BIBentrySTDinterwordspacing}{\spaceskip=0pt\relax}
\providecommand{\BIBentryALTinterwordstretchfactor}{4}
\providecommand{\BIBentryALTinterwordspacing}{\spaceskip=\fontdimen2\font plus
\BIBentryALTinterwordstretchfactor\fontdimen3\font minus
  \fontdimen4\font\relax}
\providecommand{\BIBforeignlanguage}[2]{{%
\expandafter\ifx\csname l@#1\endcsname\relax
\typeout{** WARNING: IEEEtran.bst: No hyphenation pattern has been}%
\typeout{** loaded for the language `#1'. Using the pattern for}%
\typeout{** the default language instead.}%
\else
\language=\csname l@#1\endcsname
\fi
#2}}
\providecommand{\BIBdecl}{\relax}
\BIBdecl

\bibitem{gao2020performance}
Z.~Gao, A.~Liu, and X.~Liang, ``The performance analysis of downlink {NOMA} in
  {LEO} satellite communication system,'' \emph{IEEE Access}, vol.~8, pp.
  93\,723--93\,732, 2020.

\bibitem{kodheli2020satellite}
O.~Kodheli, E.~Lagunas, N.~Maturo, Sharma \emph{et~al.}, ``Satellite
  communications in the new space era: A survey and future challenges,''
  \emph{IEEE Commun. Surv. Tutor.}, vol.~23, no.~1, pp. 70--109, 2020.

\bibitem{yin2022rate}
L.~Yin and B.~Clerckx, ``Rate-splitting multiple access for
  satellite-terrestrial integrated networks: Benefits of coordination and
  cooperation,'' \emph{IEEE Trans. Wirel. Commun.}, vol.~22, no.~1, pp.
  317--332, 2022.

\bibitem{cui2023energy}
H.~Cui, L.~Zhu, Z.~Xiao \emph{et~al.}, ``Energy-efficient {RSMA} for multigroup
  multicast and multibeam satellite communications,'' \emph{IEEE Wirel. Commun.
  Lett.}, vol.~12, no.~5, pp. 838--842, 2023.

\bibitem{khan2023rate}
W.~U. Khan, Z.~Ali, E.~Lagunas \emph{et~al.}, ``Rate splitting multiple access
  for next generation cognitive radio enabled {LEO} satellite networks,''
  \emph{IEEE Trans. Wirel. Commun.}, vol.~22, no.~11, pp. 8423--8435, 2023.

\bibitem{lee2023coordinated}
J.~Lee, J.~Lee \emph{et~al.}, ``Coordinated rate-splitting multiple access for
  integrated satellite-terrestrial networks with super-common message,''
  \emph{IEEE Trans. Veh. Technol.}, vol.~73, no.~2, pp. 2989--2994, 2023.

\bibitem{lin2020secure}
Z.~Lin, M.~Lin, B.~Champagne \emph{et~al.}, ``Secure and energy efficient
  transmission for {RSMA}-based cognitive satellite-terrestrial networks,''
  \emph{IEEE Wirel. Commun. Lett.}, vol.~10, no.~2, pp. 251--255, 2020.

\bibitem{jiang2023aerial}
C.~Jiang, C.~Zhang, L.~Mu \emph{et~al.}, ``Aerial {RIS}-aided physical layer
  security design for satellite communication among similar channels,''
  \emph{J. Intell. Inf. Syst.}, vol.~1, no.~1, pp. 54--67, 2023.

\bibitem{bao2021adep}
T.~Bao, H.~Wang, H.-C. Yang \emph{et~al.}, ``On the {ADEP} and {DOR} analysis
  of {RIS}-aided {URLLC} systems with partial {CSI} in smart factory,'' in
  \emph{IEEE GC Wkshps}, Madrid, Spain, Dec. 2021, pp. 1--6.

\bibitem{bao2022performance}
------, ``Performance analysis of {RIS}-aided communication systems over the
  sum of cascaded {Rician} fading with imperfect {CSI},'' in \emph{IEEE WCNC},
  Austin, TX, Apr. 2022, pp. 399--404.

\bibitem{lin2022refracting}
Z.~Lin, H.~Niu \emph{et~al.}, ``Refracting {RIS}-aided hybrid
  satellite-terrestrial relay networks: Joint beamforming design and
  optimization,'' \emph{IEEE Trans. Aerosp. Electron. Syst.}, vol.~58, no.~4,
  pp. 3717--3724, 2022.

\bibitem{wang2022secure}
Y.~Wang, Z.~Lin, H.~Niu \emph{et~al.}, ``Secure satellite transmission with
  active reconfigurable intelligent surface,'' \emph{IEEE Commun. Lett.},
  vol.~26, no.~12, pp. 3029--3033, 2022.

\bibitem{liu2023deep}
Y.~Liu, C.~Huang, G.~Chen \emph{et~al.}, ``Deep learning empowered trajectory
  and passive beamforming design in {UAV}-{RIS} enabled secure cognitive
  non-terrestrial networks,'' \emph{IEEE Wirel. Commun. Lett.}, vol.~13, no.~1,
  pp. 188--192, 2023.

\bibitem{zhou2024heuristic}
H.~Zhou, M.~Erol-Kantarci, Y.~Liu \emph{et~al.}, ``Heuristic algorithms for
  {RIS}-assisted wireless networks: Exploring heuristic-aided machine
  learning,'' \emph{IEEE Wirel. Commun.}, vol.~31, no.~4, pp. 106--114, 2024.

\bibitem{peng2022performance}
Z.~Peng, X.~Chen, C.~Pan \emph{et~al.}, ``Performance analysis and optimization
  for {RIS}-assisted multi-user massive {MIMO} systems with imperfect
  hardware,'' \emph{IEEE Trans. Veh. Technol.}, vol.~71, no.~11, pp.
  11\,786--11\,802, 2022.

\bibitem{schroder2023comparison}
A.~Schr{\"o}der, M.~R{\"o}per, D.~W{\"u}bben \emph{et~al.}, ``A comparison
  between {RSMA}, {SDMA}, and {OMA} in multibeam {LEO} satellite systems,'' in
  \emph{VDE WSA \& SCC}, Braunschweig, DE, Mar. 2023, pp. 1--6.

\bibitem{pei2024secrecy}
X.~Pei, Y.~Chen, M.~Wen \emph{et~al.}, ``Secrecy performance analysis of
  {RSMA}-based communications under partial {CSIT} against randomly located
  eavesdroppers,'' \emph{Early Access, IEEE Trans. Veh. Technol.}, 2024.

\bibitem{wu2024deep}
M.~Wu, K.~Guo, X.~Li \emph{et~al.}, ``Deep reinforcement learning-based energy
  efficiency optimization for {RIS}-aided integrated
  satellite-aerial-terrestrial relay networks,'' \emph{IEEE Trans. Commun.},
  vol.~72, no.~7, pp. 4163--4178, 2024.

\bibitem{ngo2023multi}
Q.~T. Ngo, B.~A. Jayawickrama \emph{et~al.}, ``Multi-agent {DRL}-based
  {RIS}-assisted spectrum sensing in cognitive satellite-terrestrial
  networks,'' \emph{IEEE Wirel. Commun.}, vol.~12, no.~12, pp. 2213--2217,
  2023.

\bibitem{wu2023joint}
M.~Wu, K.~Guo, Z.~Lin \emph{et~al.}, ``Joint optimization design of
  {RIS}-assisted hybrid {FSO} {SAGIN}s using deep reinforcement learning,''
  \emph{IEEE Trans. Veh. Technol.}, vol.~73, no.~3, pp. 3025--3040, 2023.

\end{thebibliography}

\end{document}